\documentclass[%
 reprint,
superscriptaddress,
 amsmath,amssymb,
 aps,
]{revtex4-2}

\usepackage{graphicx}
\usepackage{dcolumn}
\usepackage{bm}

\usepackage{hyperref}

\usepackage[utf8]{inputenc}
\usepackage[T1]{fontenc}
\usepackage{mathptmx}
\usepackage{etoolbox}

\makeatletter
\def\@email#1#2{%
 \endgroup
 \patchcmd{\titleblock@produce}
  {\frontmatter@RRAPformat}
  {\frontmatter@RRAPformat{\produce@RRAP{*#1\href{mailto:#2}{#2}}}\frontmatter@RRAPformat}
  {}{}
}%
\makeatother

\usepackage{hyperref}

\usepackage{commath}
\usepackage{xcolor}
\usepackage{tabu}

\newcommand{\beq}{\begin{equation}}
\newcommand{\eeq}{\end{equation}}
\newcommand{\nvec}{\hat{\bf n}}

\newcommand{\rvec}{\hat{\bf r}}

\graphicspath{ {./} }
\DeclareGraphicsExtensions{.pdf,.png}

\def\revA#1{\textcolor{black}{#1}} 
\def\revB#1{\textcolor{black}{#1}} 
\def\revC#1{\textcolor{black}{#1}} 

\begin{document}


\title{Formation and Fluctuation of Two-dimensional Dodecagonal Quasicrystal}

\author{Uyen Tu Lieu}
\email{uyen.lieu@aist.go.jp}
\affiliation{Mathematics for Advanced Materials-OIL, AIST, 2-1-1 Katahira, Aoba, 980-8577 Sendai, Japan}

\author{Natsuhiko Yoshinaga}
\email{yoshinaga@tohoku.ac.jp} 
\affiliation{Mathematics for Advanced Materials-OIL, AIST, 2-1-1 Katahira, Aoba, 980-8577 Sendai, Japan}
\affiliation{WPI-Advanced Institute for Materials Research (WPI-AIMR), Tohoku University, 2-1-1 Katahira, Aoba, 980-8577 Sendai, Japan}


\begin{abstract}
The self-assembly of two-dimensional dodecagonal quasicrystal (DDQC) from patchy particles are investigated by Brownian dynamics simulations. The patchy particle has a five-fold rotational symmetry pattern described by the spherical harmonics $Y_{55}$. From the formation of the DDQC obtained by an annealing process, we find the following mechanism. The early stage of the dynamics is dominated by hexagonal structures. Then, nucleation of dodecagonal motifs appears by particle rearrangement, and finally the motifs expand whole system. The transition from the hexagonal structure into the dodecagonal motif is made by the collective rational motion of the particles. The DDQC consists of clusters of dodecagonal motifs, which can be classified into several packing structures. By the analyses of the DDQC under fixed temperature, we find the fluctuations are characterised by changes in the network of the dodecagonal motifs. Finally we compare the DDQC assembled from the patchy particle system and isotropic particle system. The two systems both share a similar mechanism of the formation and fluctuation of DDQC.
\end{abstract}

\maketitle


\section{Introduction} 
	The quasicrystals (QCs) are ordered structures lacking periodic translational symmetry \cite{shechtman_1984,levine_1984}. Different from the crystals which possess 2-, 3-, 4-, 6-fold rotational symmetry, the quasicrystals may have 5-, 8-, 10-, 12-, 18-fold rotational symmetry. The quasicrystal can be applied in various applications, such as advanced coatings, reinforced composites, optics, photovoltaics, magnetism \cite{urgel_2016}, superconductivity \cite{kamiya_2018}, bandgap materials in photonic devices \cite{steurer_2007}. 
The appearance of quasicrystals and approximants have been observed in different length scales, from intermetallics \cite{shechtman_1984,bindi_2009} to meso-scale \cite{forster_2013,talapin_2009,mikhael_2008,dotera_2011,takagi_2017}. \revA{Recently, quasicrystalline structures have been found in soft materials, such as block copolymers \cite{lindsay_2020,hayashida_2007},  surfactants \cite{yue_2016}, colloids \cite{fischer_2011}, and dendrimers \cite{zeng_2004}}.

     
The quasiperiodic self-assembly has been obtained mainly by three types of mechanisms: isotropic interactions between particles, anisotropic interactions, or polydispersed systems. In the first case, the isotropic interaction potential depends only on the distance between particles, but has at least two length scales \cite{engel_2010,engel_2015,barkan_2014,damasceno_2017,  dzugutov_1993,  dotera_2014}. 
For example, the Lennard-Jones-Gauss potentials can produce decagonal QC, dodecagonal QC \cite{engel_2010}, or the hard-core/square-shoulder potentials \cite{dotera_2014} create a family of QC structures with 10-, 12-, 18- and 24-fold bond orientational order. Continuum description has also been proposed in ref.\cite{lifshitz_1997}. \revC{The three-well oscillating pair potential is able to form a three-dimensional icosahedral QC in non-atomic systems \cite{engel_2015}. }
	
In the second case of anisotropic interactions between particles, the interactions depend on their mutual orientation while the distance-dependence has only one length scale. For example, the five-patch particles, whose five sticky patches are equally distributed around the equator of the sphere, can arrange into a two-dimensional dodecagonal QC \cite{vanderlinden_2012,reinhardt_2013,gemeinhardt_2018,gemeinhardt_2019};  hard tetrahedra create three-dimensional dodecagonal QC \cite{je_2021}.

In the third type of systems, DDQC and approximants structures are realised in polydispersed systems \cite{talapin_2009, reinhardt_2017, liu_2019a}, such as binary nanoparticle \cite{talapin_2009}, bi-dispersed pentavalent and hexavalent patchy particles \cite{reinhardt_2017}. 




\revA{Despite the intensive studies on the structures of quasicrystals,} kinetics and dynamics of quasicrystal growth, however, are still incomplete \cite{keys_2007,steurer_2018,grimm_1999,janssen_2018}. \revA{It is not clear how the quasicrystals appear from a liquid or crystalline state, what is happening at the periphery of quasicrystals duing their growth, and how the structures fluctuate. It is also of interest whether the dynamics and kinetics of quasicrystals are dependent or not on the three aforementioned} mechanisms. Most of the previous studies focused on the analysis of the structures. In ref. \cite{vanderlinden_2012,reinhardt_2013}, stable DDQC was generated via Monte-Carlo simulations of the five-patch particles. In ref. \cite{gemeinhardt_2018}, the growth of DDQC was studied by particle deposition on a prepared quasicrystalline substrate. The growth of a three-dimensional DDQC was studied in ref. \cite{keys_2007}, and it was suggested that characteristic structures (icosahedrons in that system) preferentially appear around the nucleus of DDQC. The growth of a two-dimensional DDQC was also studied in ref. \cite{achim_2014}. Nevertheless, what kind of local structures preferentially appear near the nucleus of quasicrystals in other systems remains unknown. Also, it is of relevance to clarify how those structures are transformed as a part of DDQC by particle displacements. 

In this study, we investigate the kinetics and dynamics of DDQC by computer simulations. \revA{The majority of reported quasicrystals in soft materials are DDQC. Therefore, we focus on the formation of DDQC.} We have developed a model for the interaction of a pair of anisotropic particle whose surface pattern is described by spherical harmonics $Y_{lm}$ \cite{lieu_2020, lieu_2022}. Using the five-fold symmetry patchy particles, we are able to assemble the dodecagonal quasicrystal. We investigate the formation of the DDQC from a bulk phase, and the fluctuation after the DDQC is formed. The same procedure is applied to the DDQC assembled from the isotropic particle system. We compare the kinetics and dynamics of the DDQCs from the patchy particle system and isotropic particle system.

\section{Methods} 
\label{methods}
\subsection{Numerical simulation} 
\begin{figure}[h]  	
	   \centering
	   \includegraphics[width=0.40\textwidth, trim=15mm 205mm 100mm 30mm,clip]{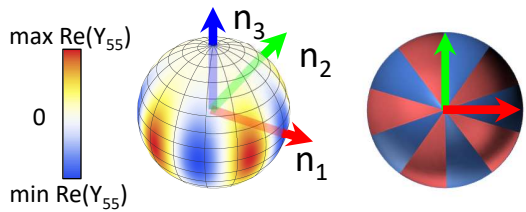}
	   \caption{Illustration of a patchy particle with five-fold symmetry expressed by the pattern of $Y_{55}$ \revC{and the corresponding patchiness}. The patches in the same colours are attractive, while those in the different colours are repulsive.}
	   \label{fig:patchy}
	\end{figure}

We consider the patchy particle with five-fold symmetry, composing of ten alternating patches of two different types (Fig.~\ref{fig:patchy}). The pattern on a spherical particle is described by spherical harmonics $Y_{55}$ \cite{lieu_2020,lieu_2022}.	
	
	The pairwise interaction of the particle include a Weeks-Chandler-Andersen (WCA) term preventing the overlap and a Morse-like, orientation-dependent term. The details of the anisotropic potential is given in Appendix \ref{app:potential}. Figure \ref{fig:potential} illustrates the potential used in the study at \revA{different} particle configurations.
	\begin{figure}[h]  
	   \centering
  		\includegraphics[width=0.45\textwidth, trim=35mm 83mm 38mm 86mm,clip]{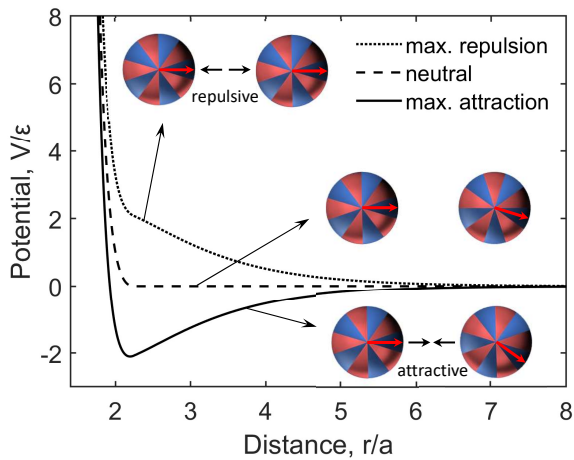}
	   \caption{Pairwise potential of patchy particles at \revA{different configurations. The red arrows indicate the orientations of the patchy particles.}}
	   \label{fig:potential}
	\end{figure}
	
	The assembly of patchy particles is performed by Brownian dynamics simulations \cite{allen_2017}. 
	\revC{The positions $\mathbf{r}$ and orientations $\bm{\Omega}$ of the particles after the time step $\Delta t$ are updated according to the equations:	
	\begin{equation}
	 \mathbf{r}(t+\Delta t)
	 = \mathbf{r}(t) 
	 + \frac{D^\text{T}}{k_\text{B} T} \mathbf{F}(t) \Delta t 
	 + \bm{\delta}^\text{G} \sqrt{2 D^\text{T}\Delta t}, 
	 \end{equation}
	 \begin{equation}
	 \bm{\Omega}(t+\Delta t)
	 = \bm{\Omega}(t) 
	 + \frac{D^\text{R}}{k_\text{B} T}  \mathbf{T}(t) \Delta t 
	 + \bm{\delta}^\text{G} \sqrt{2D^\text{R} \Delta t},
	 \end{equation}
where $D^\text{T},D^\text{R}$ are the translational and rotational diffusion coefficients, respectively; the force $\mathbf{F}$ and torque $\mathbf{T}$ are derived from the pair potential $V$; and each component of $\bm{\delta}^\text{G}$ is a Gaussian distribution with zero mean and unit variance. The simulations are conducted in dimensionless form where the characteristic length, energy, time, and temperature, are the particle radius $a$, Lennard-Jones potential well-depth $\epsilon$, the Brownian diffusion time $\tau_\text{B}=a^2/D^\text{T}$, and $\epsilon/k_B$, respectivetly}.   
  
	The spherical particles are confined to a two-dimensional plane while rotate freely in three dimensions. \revA{The initial condition of the simulation is random distribution of positions in a periodic box of the size $L_x \times L_y$ with random orientations}. The number of particles is $N=1024$. \revC{The density is defined as area fraction $\rho_\text{a}=\pi N/(L_x L_y)$.}
	
		We consider the systems undergoing annealing and at fixed temperature. In the annealing, the temperature is varied from $T_{max}=1.2$ to $T_{min}=0.4$ with intervals of $\Delta T=0.0125$. At each temperature the number of steps is $0.1\times10^6$, and the total steps are $6.5 \times 10^6$. \revB{The time step is set at $\Delta t=0.5\times10^{-3}T$, thus the total time of the annealing simulation is $t=2600$.} \revA{We have checked the annealing schedule is slow enough to make stable quasicrystals.} For the system at fixed temperature, the system at the initial condition is suddenly quenched at the temperature $T$, and is relaxed at the fixed $T$. We conduct simulations at fixed temperature $T \in [0.5, 1.1]$, the number of simulation steps is $5 \times 10^6$, \revB{the simulation time at each temperature is $t=2500T$.}
	
	\revB{To analyse the fluctuation of the DDQC, simulations are performed with the parameters: area fraction $\rho_{a}=0.75$, fixed temperature $T=0.8$ and $50\times 10^6$ steps, which are approximately 10 times longer and slightly less dense than the annealing simulation. For five independent simulations, the DDQC is formed from a rich hexagonal phase after the first $5 \times 10^6$ steps. The fluctuation of the DDQC is reported after it is formed.} 	
	\revB{We also investigate the DDQC assembled from the isotropic interacting particles. The details of the potential and simulation condition are given in Appendix \ref{app:isotropic}.}

\subsection{Structural analysis} 
\begin{figure}[ht] 
   	\centering   	
   	\includegraphics[width=0.48\textwidth, trim=5mm 92mm 10mm 92mm,clip]{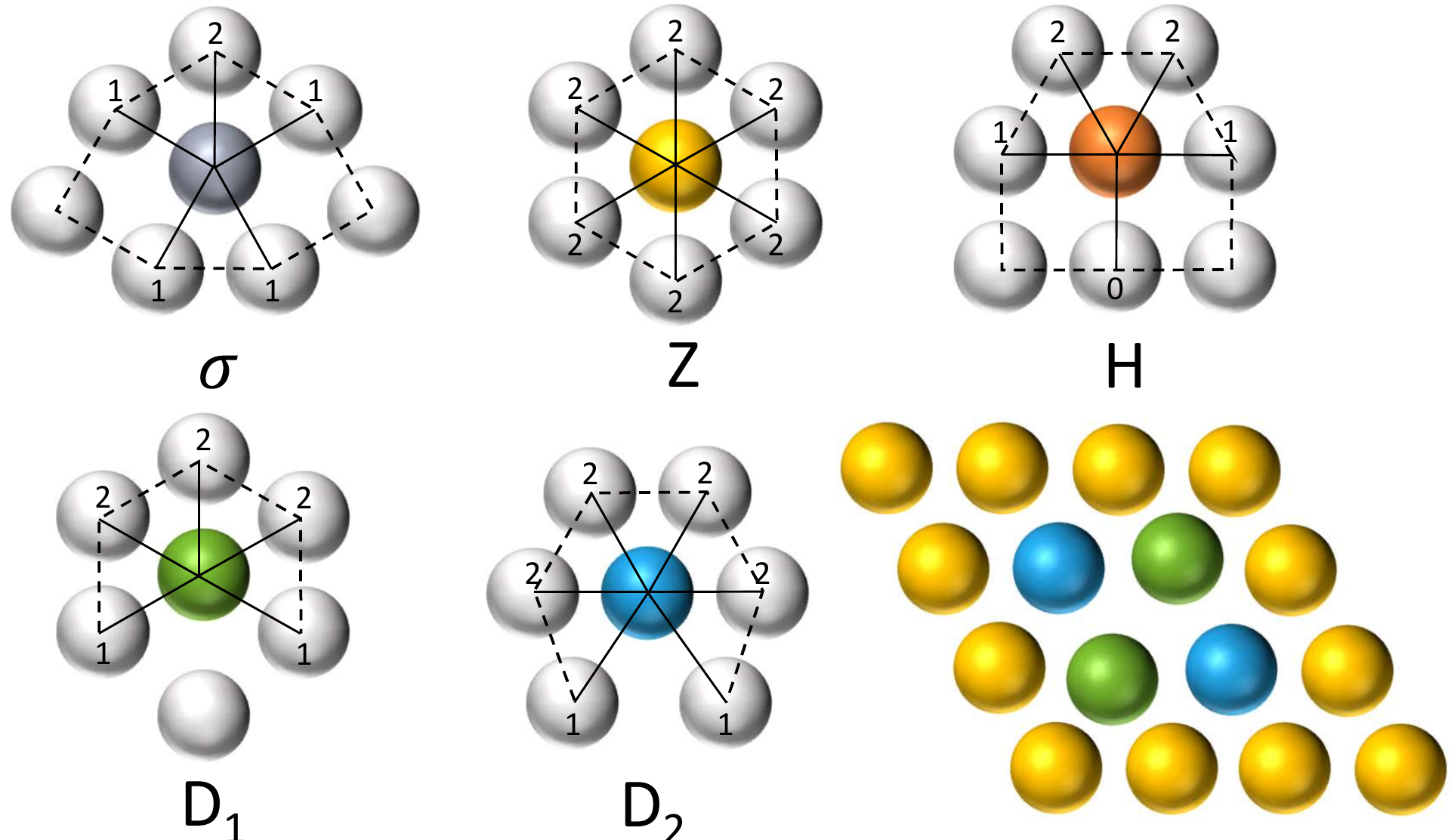}
   	\caption{Local structure $\sigma$, $Z$, $H$, $D_1$, $D_2$ and its nearest neighbours $r \leq 2.5a$, and the illustration of the diamond structure $D_1$, $D_2$ inside a cluster of $Z$ particles. The common neighbours are also given.}
   	\label{fig:local.structure}
	\end{figure}

The types of local structure of each particle are determined by the connection of the nearest neighbours of a particle. The local environments in a dodecagonal quasicrystal are $\sigma$, $H$, $Z$ \cite{vanderlinden_2012} as given in Fig.~\ref{fig:local.structure}. These local structures are based on the number of common neighbours of the neighbouring particles analogial to the Frank-Kasper phases. \revA{For example, a $\sigma$ particle has five neighbours. One of them has two common neighbours, and four of them have only one common neighbour. Therefore the local structure of a $\sigma$ particle is identified as $\{21111\}$. The other local structures are $H$ $\{$22110$\}$, $Z$ $\{$22222$\}$, $D_1 $ $\{$22211$\}$, $D_2$ $\{$222211$\}$.} In this study, the new local structures $D_1$, $D_2$ result from fluctuating $Z$ particles, where they form a diamond shape shown in Fig.~\ref{fig:local.structure}. The particle which does not fall into these categories is considered as undefined $U$. \revA{The number of unidentified particles is only $\lesssim 5\%$ of the total number of the particles in the system.}

The local structure of the particle varies with time. The interchange of a given local structure to/from other local structures is determined between two \revA{two consecutive snapshots different by $10^4$ time steps}, for each local structure. \revB{The purpose of this calculation is to identify the relevant ``reaction'' of the local structures during the growth of DDQC. For example, we can calculate the interchange rate between $Z$ and $\sigma$. If the rate of $Z \rightarrow \sigma$ is higher than that of $\sigma \rightarrow Z$, then the speed of $Z \rightarrow \sigma$ is faster than $\sigma \rightarrow Z$.}

\begin{figure}[ht] 
  	\centering
   	\includegraphics[width=0.45\textwidth]{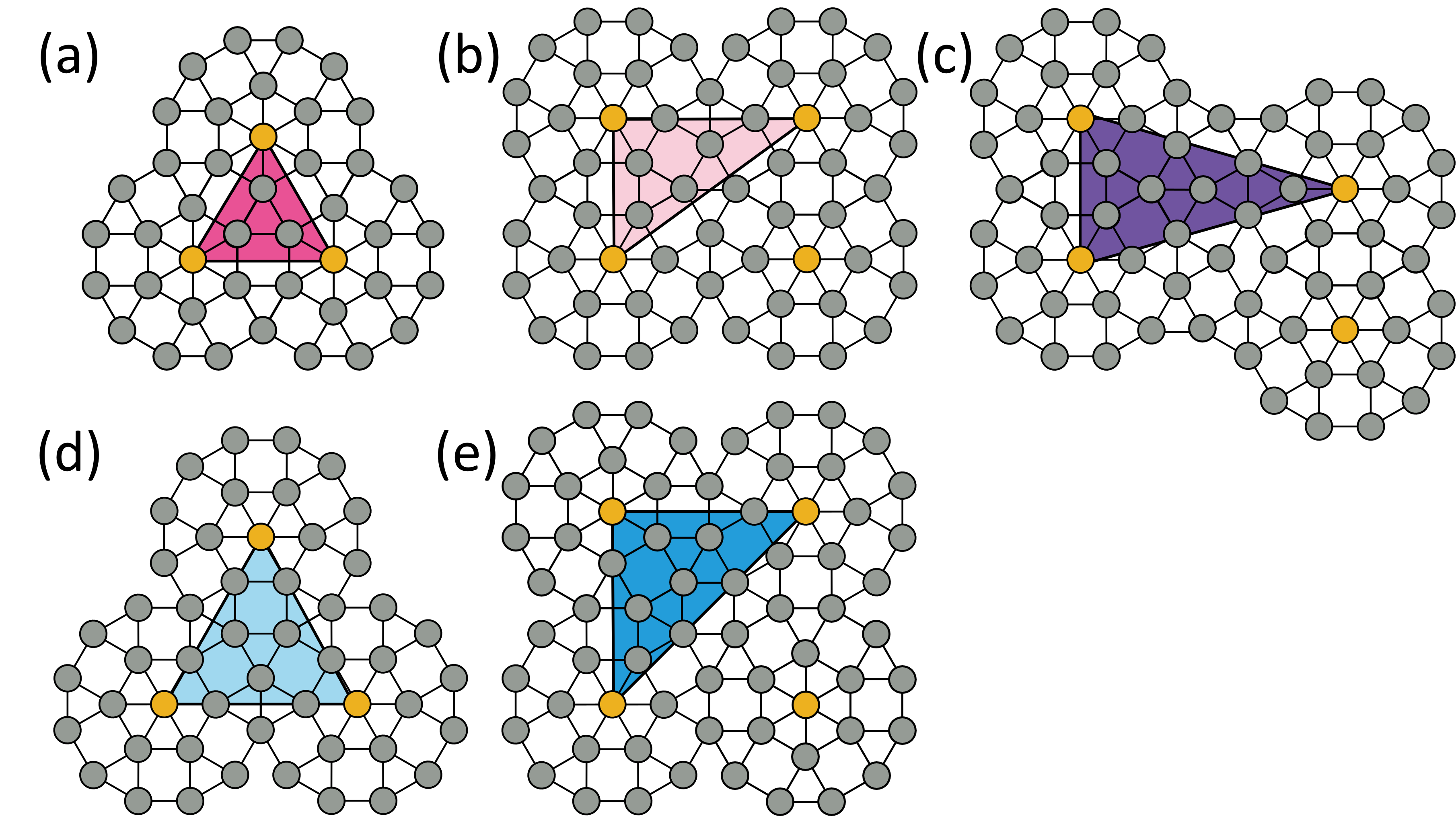}   	
   	\caption{Typical packing of dodecagonal motifs. \revB{The packing is characterised by the Delaunay triangulation of the dodecagonal motif centres. Particle colours are given in Fig.~\ref{fig:local.structure}. } }    
   	\label{fig:motif.pack}
	\end{figure}

\begin{figure*}[ht] 
   	\centering 	
 	\includegraphics[width=0.9\textwidth, trim=10mm 145mm 10mm 10mm,clip]{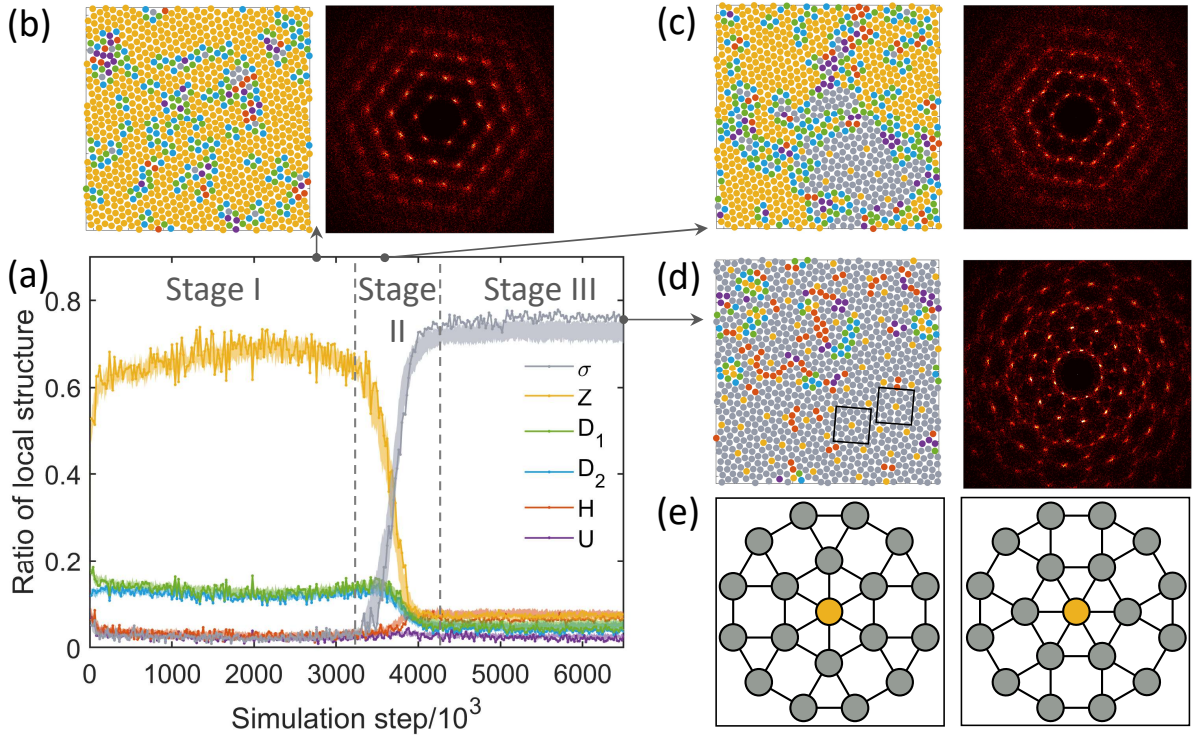}        
   	\caption{Change in local structures during annealing simulations at $\rho_\text{a}=0.78$. (a) Ratio of the local structure \revA{in time}; the line represents the data of a simulation, the shade area corresponds to the 95$\%$ confidence intervals of 10 independent simulations. The growth of the dodecagonal quasicrystal \revA{is divided into three stages.} (b-d) The snapshots and their Fourier transformation; the colour of particle indicates the local structure as shown in the legend and Fig.~\ref{fig:local.structure}. (e) Illustration of two kinds of dodecagonal motifs in the boxes in (d), they are interchangeable by rotating the particles on the first ring $\pm30^\circ$.}
   	\label{fig:snapshot}
	\end{figure*}

The DDQC has a characteristic dodecagonal motif \cite{reinhardt_2013,forster_2013} (see Fig.~\ref{fig:snapshot}(e)). The motif is made of one $Z$ particle at the centre, six $\sigma$ particles on the first ring, and twelve $\sigma$ particles on the second ring. Figure \ref{fig:motif.pack} shows several ways of packing the motifs. \revA{These dodecagonal motifs are identified by the network of $Z$ particles in each snapshot of simulations. The $Z$ particles are centres of each dodecagonal ring. We describe the motif packing by the triangular tiles in the network of $Z$ particles obtained from Delaunay triangulation}. \revC{Different clusters of motifs are shown in different colours. Each cluster is characterised by the distance between the dodecagonal motifs. For example, the dodecagonal motifs can interpenetrate in Fig.~\ref{fig:motif.pack}(a), share a common edge in Fig.~\ref{fig:motif.pack}(d-e); or connect via extra $\sigma$ particles in Fig.~\ref{fig:motif.pack}(e). The criteria of the clusters are based on the edge lengths of the triangular tiles; e.g. the edge lengths of the perfect cluster in Fig.~\ref{fig:motif.pack}(a) are $l=(1+\sqrt{3})2a$; a simulated cluster is of this type if the edge lengths $l_\text{sim}$ satisfy $|l_\text{sim} - l| < 0.1l$. We will discuss how the structure of the clusters of motifs changes in time. The time-dependent network structure during $[t, t+\delta t]$ is visualised by overlapping the images at $[t,\ldots,t+\delta t]$ (see Fig.~\ref{fig:fluctuation2}, \ref{fig:isotropic.rotation}). }   
	



\section{Results of patchy particle system}
\subsection{The formation of dodecagonal quasicrystal}
\label{sec:31}
The dynamical self-assembly using the \revA{five-fold symmetric} patchy particles under annealing is investigated. Figure \ref{fig:snapshot} shows the  change in the ratios of local structures during the formation of DDQC and the representative snapshots with \revA{the colour coded according to local structures} $\sigma$, $Z$, $D_1$, $D_2$ (Fig.~\ref{fig:local.structure}). \revA{Dynamical formation of a dodecagonal quasicrystal can be decomposed into three stages. At stage I,} the majority of particles are the $Z$ type, with a ratio of around 0.7. \revA{Both the ratios of $D_1$ and $D_2$} are around 0.1. The snapshot of this stage is a $Z$-rich structure with clear hexagonal spots in its Fourier transform. The $D_1$ and $D_2$ particles often appear in pairs as a diamond shape surrounding by $Z$ particles. 
 Stage II corresponds to the growth of the QC where the ratios of $Z$, $D_1$, $D_2$ decrease, while that of $\sigma$ increases. A $\sigma$-rich region \revA{is formed in the bulk of $Z$ particles. The $\sigma$-rich region contains a few $Z$ particles inside.
 The $D_1$ and $D_2$ particles accumulate at the boundary between this region and the `bulk' $Z$ particles.} 
Finally at stage III, when \revA{the expansion of $\sigma$ reaches the system size, all the ratios maintain constant values with fluctuations.} The $\sigma$-rich region expands \revA{whole in the system} and a dodecagonal quasicrystal (DDQC) is assembled, as we can observe the twelve-fold rotational symmetry in the Fourier space \cite{vanderlinden_2012, forster_2013}. 
 
  	There are several dodecagonal motifs in the QC structure. The motif contains one $Z$ particle at the centre and six $\sigma$ particles on the first ring and twelved $\sigma$ particles on the second ring (see Fig.~\ref{fig:snapshot}(e)). There are two types of motifs which are different by 30$^\circ$ degrees of the first ring. The QC structures consists of the two types of dodecagonal motifs packed in different ways as shown in Fig.~\ref{fig:motif.pack}. 
  	\revA{The four structures in which the $Z$ particles of the motif centres form either \revB{equillateral triangle (Fig.~\ref{fig:motif.pack}(a,d)) or right angle triangle (Fig.~\ref{fig:motif.pack}(b,e)),} were obtained in \cite{vanderlinden_2012} using patchy particles with five-fold or seven-fold symmetry. In our simulations, we also obtain the structure in which the $Z$ particles form} \revB{isosceles triangle} (Fig.~\ref{fig:motif.pack}(c)). It is noted that the term ``dodecagonal quasicrystal'' here is loosely used for a structure whose Fourier transform has spots with clear twelve-fold symmetry. The obtained structures are not clear enough to be identified as quasicrystal or approximants. The structure contains both defects and ordered dodecagonal motifs.   		
	\begin{figure}[ht] 
   	\centering 	    
 	\includegraphics[width=0.45\textwidth, trim=10mm 135mm 105mm 5mm,clip]{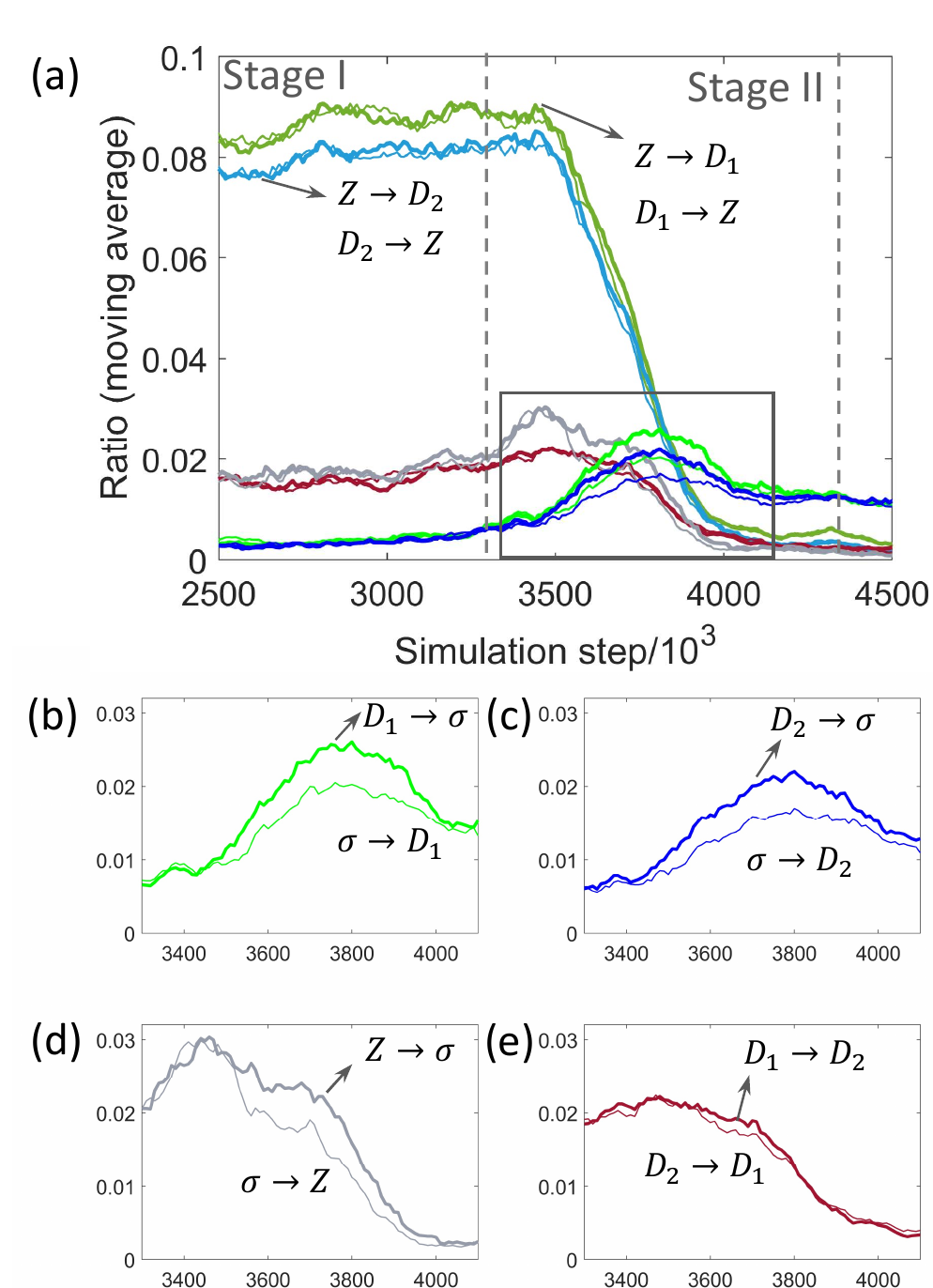}
   	\caption{Interchanges between the local structures during the growth of DDQC at $\rho_\text{a}=0.78$ by decomposition of the ``in'' and ``out'' amount of selected pairs of local structures (a). (b-e) The corresponding data inside the box in (a). The graph show moving average of 15 data points.}
   	\label{fig:interchange}
	\end{figure}	

The ratio in Fig.~\ref{fig:snapshot}(a), however, can not show the ``reaction'' rate of each local structure change to/from other ones.  
Figure~\ref{fig:interchange} shows the interchange among $Z,D_1,D_2,\sigma$. During stage I, a majority of $Z$ particles changes to and from $D_1, D_2$ as the ``reactions'' are comparable in both forward and backward directions. In stage II, the particles $Z, D_1, D_2$ change to $\sigma$ faster than the backward direction. It suggests that the \revB{relevant ingredients for} the formation of the $\sigma$ particle, or the dodecagonal motif, are the $Z, D_1, D_2$ particles.  
  	
\begin{figure}[ht] 
  	\centering
	\includegraphics[width=0.47\textwidth, trim=12mm 127mm 14mm 5mm,clip]{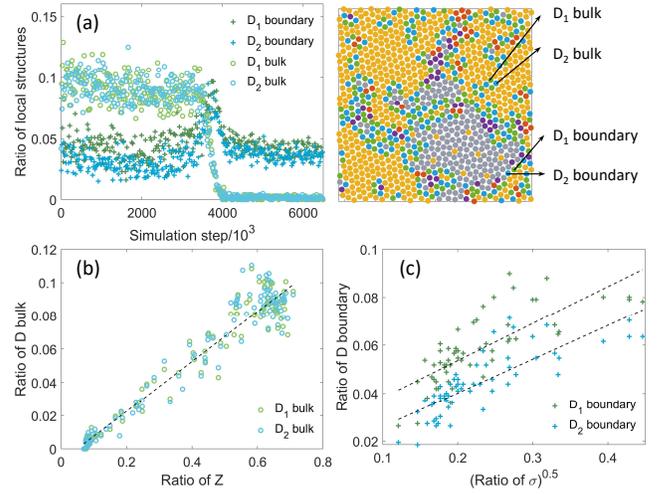}  
   	\caption{Decomposition of the local structure $D_1$, $D_2$ into the corresponding $D_1$, $D_2$ in bulk and in boundary. (a) The composition at bulk and boundary; the illustration shows the bulk particles and the boundary particles. (b) Scaling of bulk $D$ with $Z$ for simulation steps \revA{$[3000\times 10^3,4100 \times 10^3]$. (c) Scaling of boundary $D$ with $\sigma$ for simulation steps $[3000\times 10^3,3600 \times 10^3]$.} The dashed lines are regression data.}
   	\label{fig:D}
	\end{figure}

We propose that there is a relation between $Z,D_1,D_2$, and $\sigma$. Specifically, the particles $D_1$, $D_2$ play as an intermediate role for the formation and expansion of the dodecagonal quasicrystal from a $Z$-rich structure. Figure \ref{fig:D}(a) shows the decomposition of the $D_1$, $D_2$ particles in terms of the bulk $D_{\text{bulk}}$ and the boundary $D_{\text{boundary}}$ amounts. The $D_{\text{boundary}}$ are defined as the $D$ particles ($D_1$ and $D_2$) neighbouring to the $\sigma$ particles, while the rest is defined as $D_{\text{bulk}}$. The bulk $D$ fluctuates around a constant value, then decreases to zero as the QC occupies all the space. Figure \ref{fig:D}(b) show the relation between the bulk $D$ particles and $Z$ particles. For the bulk particles, the amount of $D_1$ and $D_2$ is almost same as they often appear as pairs (see snapshots in Fig.~\ref{fig:snapshot},~\ref{fig:D}), and linearly increase with $Z$. \revA{We fit the data in Fig.~\ref{fig:D}(b) by linear regression, and find} the slope of $6.6^{-1}$. In Fig. \ref{fig:local.structure} we illustrate a diamond shape made of four $D_\text{bulk}$ in twelve $Z$. In this case, the ratios of $D_1$ and $D_2$ to $Z$ is equal as $6^{-1}$. \revA{This argument supports a picture in which $D_1$ and $D_2$ particles appear as fluctuations of $Z$ particles and appear in pairs.} 

Regarding the $D$ particles at the boundary, as given in Fig.~\ref{fig:D}(a), these particles occupy a small amount before the growth \revA{of the $\sigma$-rich domain}, then increases during the growth, and eventually decreases to a plateau. 
Before the QC is formed, the number of stable $D$ particles is small and they appear stochastically associated with $\sigma$ particles; then when the QC is expanded, the number of boundary $D$ particles increases. The decreases of the boundary $D$ is because of the limit of the \revA{finite system size}. Moreover, the number of $D_1$ particles is higher than $D_2$ particles, suggesting that there is an extra $D_1$ which may contribute to the transformation from $D_1$ to $\sigma$. Figure \ref{fig:D}(c) shows that the $D_\text{boundary }$ grows linearly with $\sigma^{0.5}$. This is because the growing QC has the area $\propto N_\sigma$ and the circumference $\propto \sqrt{N_{\sigma}}$.  

\begin{figure}[ht] 
   \centering
  	\includegraphics[width=0.45\textwidth, trim=12mm 85mm 55mm 5mm,clip]{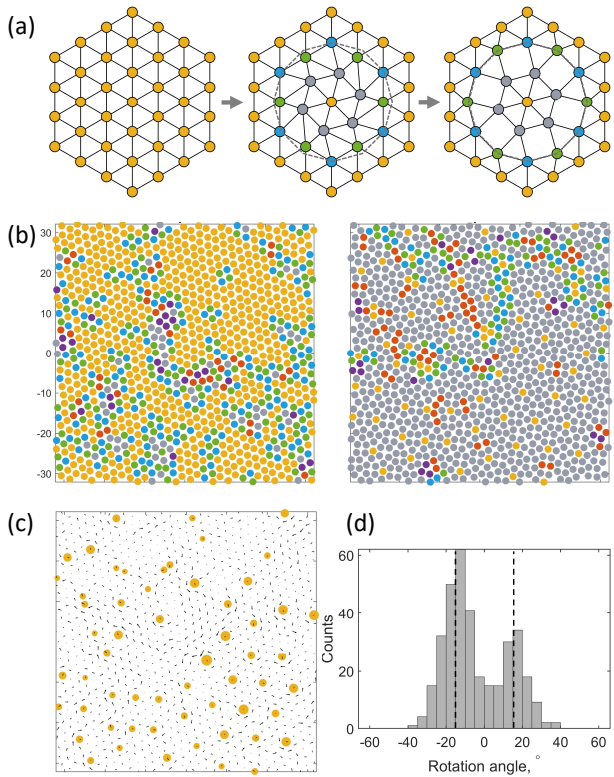}  
   \caption{Local rotation during the growth of a DDQC. (a) Illustration of local rotation from a hexagonal structure to a dodecagonal motif by rotating the particles on the first ring in clockwise direction $15^\circ$. (b) Two snapshots before and after the DDQC is formed. (c) Corresponding displacement for the snapshots in (b). \revC{The yellow dots are the $Z$ particles in the DDQC.} The size of $Z$ particles indicates the ``strength'' of the rotation of the first ring. \revC{The $Z$ particles of very low rotation are not shown.} The displacement of the first ring of the $Z$ particle reveals a rotation. \revA{(d) Histogram of the rotations of the neighbouring particles of the $Z$ particles in (c). The vertical dashed lines show rotations at $\pm 15^\circ$.}
   }
   \label{fig:mechanism}
   \end{figure}

We seek for a simple mechanism of DDQC growth. In the simulation, the particles fluctuate locally and change their local structures continuously. However, we observe a typical local rotation of the particle during growth of DDQC. Figure \ref{fig:mechanism} suggests a mechanism from a $Z$-rich structure (stage I) to a DDQC (stage III). In the schematic of Fig.~\ref{fig:mechanism}(a), from a hexagonal lattice, the first ring of the hexagonal lattice rotates by 15$^{\circ}$ either clockwise or anticlockwise. Simultaneously some particles of the second ring displace and eventually a dodecagonal motif is formed. The $D_1$, $D_2$ particles also appear during this local rotation, and become $\sigma$ or $Z$ as the dodecagonal motif grows. We can calculate the rotation of the first neighbours of a centre particle (see Appendix \ref{app:rotation}) from the displacements of the particles between two snapshots in Fig.~\ref{fig:mechanism}(b). Figure Figure \ref{fig:mechanism}(c) shows that around the centres of the DDQC motif (roughly the $Z$ particles in the DDQC), there are rotations for the particles on the first rings. The histogram in Fig.~\ref{fig:mechanism}(d) approximately has two peaks at $\pm 15^{\circ}$.

\revA{To summarise, }the mechanism of the self-assembled DDQC from patchy particles can be proposed as follows: (i) initially the structure is rich in $Z$ particle and thermal fluctuation induces the pairs of $D_1$ and $D_2$ \revA{in the bulk of} $Z$, (ii) at the right temperature, the $\sigma$ particles are formed and they cluster together to dodecagonal motifs. (iii) the DDQC rapidly expands at the boundary where the transformation $Z \rightarrow (D_1, D_2) \rightarrow \sigma$ occurs. Around the centre of the dodecagonal motifs, local rotations of the nearest neighbours take place.

\subsection{Temperature dependent}
\begin{figure}[ht] 
   \centering   
   \includegraphics[width=0.40\textwidth, trim=30mm 30mm 40mm 5mm,clip]{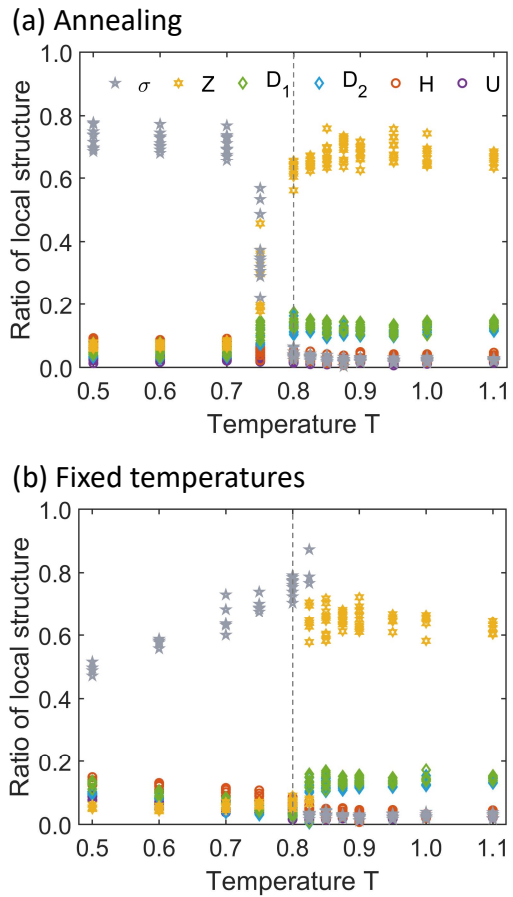}  
   \caption{Dependence of ratio of local structure on temperature in (a) annealing simulations, (b) fixed temperature simulations \revC{at $\rho_{\text{a}}=0.78$}. The dashed lines estimate the critical temperature $T^{*}$.}
   \label{fig:ratio.vs.temp}
   \end{figure}

\revA{In this section,} the effect of temperature on the DDQC is investigated. Figure \ref{fig:ratio.vs.temp} shows the dependence of local structure ratio on temperature for annealing simulations and fixed temperature simulations. In both cases, there are two phases: $Z$-rich and $\sigma$-rich, separated by a critical temperature $T^*\approx 0.8$. The $Z$-rich phase corresponds to $T > T^*$, where the structure is dominated by $Z$ particles and some $D_1$, $D_2$ particles due to the strong thermal fluctuation. \revA{In this case, the global structure is hexagonal.} The \revA{$\sigma$-rich region appears at} $T < T^*$, where the structure is dominated by $\sigma$ particles. There is a slight difference in $T^*$ of the annealing case and fixed temperature case. This is expected because the onset of the quasicrystal, i.e. the growth from the first dodecagonal motif, depends on both temperature and time, and these two parameters can not be directly compared in the two simulation procedures. For example, at $T\approx T^*$, the onset of annealing QC is in a uniform manner, whereas it is more scattered for fixed temperature QC. As a result, among ten independent simulations at fixed temperature $T=0.825$, three of them are DDQC, and the remaining seven are $Z$-rich. Regarding the quality of the QC of fixed temperature simulation, the ratio of $\sigma$ particle decreases as $T$ decreases. This is because the simulation is equivalent to very fast \revA{quenching before fixing the temperature}, the particles are kinetically trapped; hence the structure is more disordered and contains more defects. Such a behaviour is not observed in the slow annealing case, as the ratio of $\sigma$ is independent of temperature. The results reveal that the patchy particle DDQC can be obtained by annealing \revA{easier than fixed temperature} when $T\leq T^*$. The self-assemblies of high quality DDQC corresponds to the annealing scheme or fixed temperature at $T=T^*$. 
   
\subsection{Local fluctuation after DDQC is formed} 
	\begin{figure}[ht] 
  	\centering
   \includegraphics[width=0.45\textwidth, trim=35mm 54mm 25mm 7mm,clip]{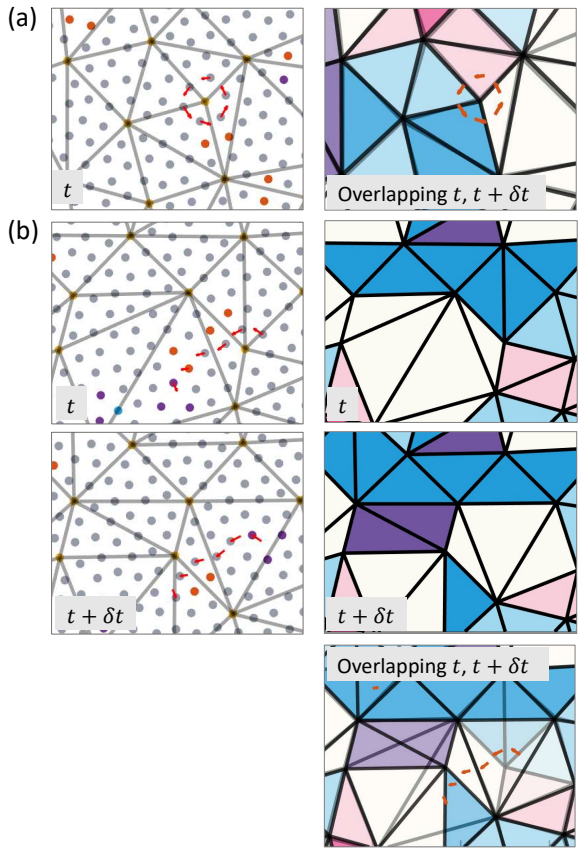}
   	\caption{Example of fluctuation of dodecagonal motif network with $\delta t=10^5$ steps at $T=0.8$ when the network (a) locally maintains or (b) changes. Left panels: network of $Z$ particles and the displacements of the particles between two snapshots. Right panels: illustration of the network in the left; colours corresponding to different packed dodecagonal motifs in Fig.~\ref{fig:motif.pack}, the white triangles are ill-defined motif packings. The coloured dots represent particle positions and particle types like Fig.~\ref{fig:local.structure}. The red arrows shows the displacements of the particles with $|r(t+\delta t)-r(t)| \geq 0.5a$. In the overlapping snapshots in (b), bold and clear edges/faces suggest the structure maintains, whereas blur ones indicate a temporal change of structure.}
   	\label{fig:fluctuation1}
	\end{figure} 
			
	\begin{figure}[ht] 
  	\centering
  	\includegraphics[width=0.45\textwidth, trim=19mm 32mm 60mm 72mm,clip]{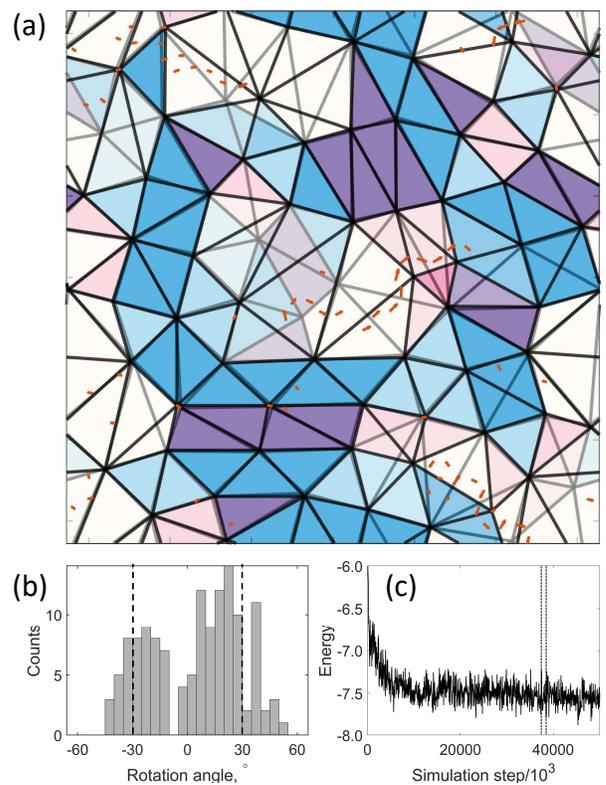}
   	\caption{Fluctuation of dodecagonal motif network after a long time scale $\delta t=10^6$ steps at $T=0.8$. (a) Collective movement of the particle as rearrangement of the network. (b) Histogram of the rotation angle of the particles in (a) with peaks at $\pm30^\circ$ (dashed lines). (c) Energy of the whole simulation process, the displacement is calculated for the time steps marked by the vertical dot lines. }
   	\label{fig:fluctuation2}
	\end{figure} 
After the DDQC is formed, we have observed that there \revA{are still structural fluctuation of the DDQC when it is subjected to thermal fluctuation}.  The QC structures consist of two types of dodecagonal motifs packed in different ways shown in Fig.~\ref{fig:motif.pack}. 
We focus on \revA{a network of the dodecagonal motifs, which is evaluated from} the network of the $Z$ particles
 \revB{because the $Z$ particles are the centres of the dodecagonal motifs.} Here we illustrate the fluctuation of the network in a short and a  long timescales. 
 
 Figure \ref{fig:fluctuation1} illustrates the $Z$ particle network by Delaunay triangulation \revB{of two snapshots within a ``short'' time scale of $\delta t =10^5$ steps. } Two cases are observed: (i) the network does not change regardless the local displacement (Fig.~\ref{fig:fluctuation1}(a)). In the dodecagonal motif, the whole first ring of the centre $Z$ can rotate 30$^{\circ}$ and eventually the dodecagonal motif is still maintained (see Fig.~\ref{fig:snapshot} for the two motifs). (ii) As shown in Fig.~\ref{fig:fluctuation1}(b), a few particles on the first ring rotate around $Z$, and cause changes in the network of the dodecagonal motif. In both cases, the displacements of the particles mostly occur for the first neighbours of the $Z$ particle \revA{as shown in Fig.\ref{fig:fluctuation1}. The displacement may occur as a chain in Fig.\ref{fig:fluctuation1}. In this case, the rearrangement of the network is coincident with the displacement.} 
 
\revA{In Figure~\ref{fig:fluctuation2}, we show fluctuations in a longer time scale ($\delta t =10^6$ steps)} \revC{by means of the packing of motifs for the whole QC structure}. \revA{The darker colours indicate that the network of $Z$ particles is maintained. The lighter regions indicate network rearrangement. There} is a collective motion of the particles along the place where \revA{rearrangement of the network of the dodecagonal motifs occurs. Readers consult Fig.~\ref{fig:isotropic.rotation} for clearer visualisation of the maintained network and network rearrangement. We perform the same analysis of rotational displacement in Sec.~\ref{sec:31}.} The rotation angles of those particles are around $\pm30^\circ$ as depicted in Fig.~\ref{fig:fluctuation2}(b). The structural change of the dodecagonal network is associated with the local rotation of the first neighbours. \revC{Such rotations seem to appear at the border of different types of the packing of dodecagonal motifs.} 
\revA{During the structural changes associated with network rearrangement of dodecagonal motifs, the energy is fluctuating around the constant value as shown in Fig.~\ref{fig:fluctuation2}(c).}

\section{Results of isotropic particle system}
\revA{As mentioned in Introduction,} the DDQC can be assembled also from isotropic particles. We perform same analysis to the isotropic DDQC. The details of the potential and simulation parameters can be found in Appendix \ref{app:isotropic}. 

\begin{figure}[ht] 
  	\centering
  	\includegraphics[width=0.50\textwidth, trim=15mm 110mm 95mm 25mm,clip]{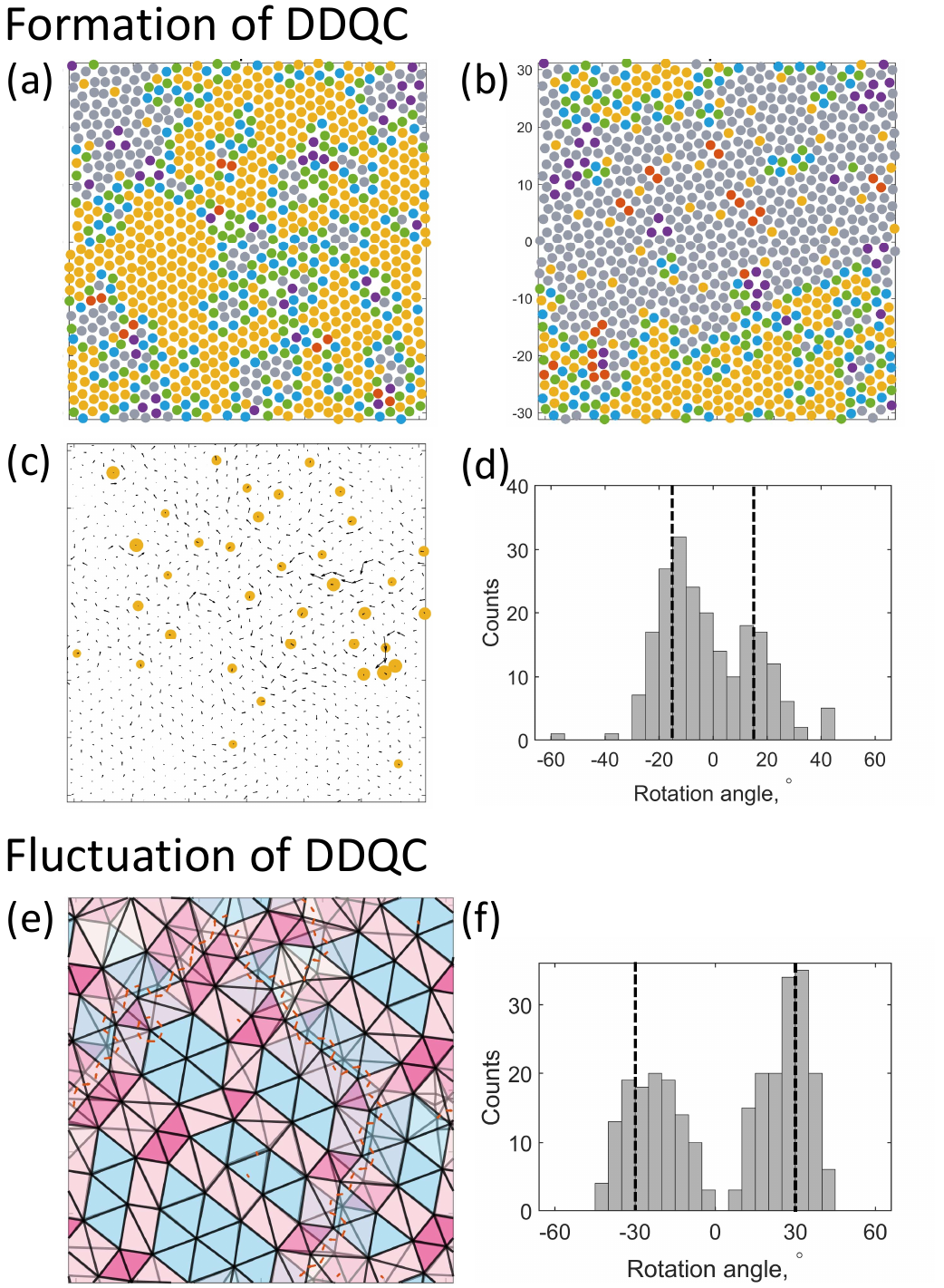}
   	\caption{\revC{Typical rotations of particles in the (a-d) formation and (e-f) fluctuation of the DDQCs. (a-b) The two snapshots for the growth of the DDQC. (c) Corresponding particle displacement of the two snapshots in (a-b). The yellow dots are the $Z$ particles in (b). The size of $Z$ particles indicates the ``strength'' of the rotation of the first ring. The $Z$ particles of very low rotation are not shown. (d) Histogram of the rotations of the neighbouring particles of the $Z$ particles in (c); the vertical dashed lines show rotations at $\pm 15^\circ$. (e) Change in network of the dodecagonal motifs during the fluctuation of a DDQC. (f) Histogram of the rotation of the particles in (e); the vertical dashed lines show rotations at $\pm 30^\circ$.}}
   	\label{fig:isotropic.rotation}
	\end{figure}

\revB{The formation and fluctuation of the isotropic DDQC is presented in Fig.~\ref{fig:isotropic.rotation}(a-b). Similar to the formation of the annealed patchy particle DDQC, the isotropic DDQC also start with a $Z$-rich structure with some $D_1, D_2$ particles generated due to fluctuation of $Z$. The dodecagonal motif is then organised and expanded, in which the $D_1,D_2$ are accumulated at the boundary of the cluster of dodecagonal motif and the $Z$-rich bulk. The resulted DDQC consists of dodecagonal motifs packed in different ways (Fig.~\ref{fig:motif.pack}). Regarding the mechanism of the transformation from $Z$-rich structures to DDQCs, the first neighbours of the dodecagonal motif centres usually rotate by $\pm 15^\circ$ (Fig.~\ref{fig:isotropic.rotation}(c-d)). The fluctuation of the DDQC is investigated by performing simulations at fixed temperature at $T=0.7$}. \revB{Figure \ref{fig:isotropic.rotation}(e) illustrates the fluctuations of isotropic DDQC in terms of the network of the dodecagonal motifs. The change in the network is also associated with the collective rotation by $\pm 30^\circ$ of the first neighbours of the motif centres.} \revC{For a detailed comparison with Fig.~\ref{fig:snapshot}, \ref{fig:interchange}, \ref{fig:ratio.vs.temp} in patchy particle system, the corresponding analyses for isotropic particle system are given in Fig. S1-S3 in Supplementary Information.} 

\revC{However, there are differences between the DDQCs made of patchy particles and isotropic particles. In general, the isotropic DDQC (Fig.~\ref{fig:isotropic.rotation}(e)) has less defects as the network has more well-defined packing structures than that of patchy particle DDQC in Fig.~\ref{fig:fluctuation2}.} 

\revC{In a dodecagonal motif of the DDQC, the $Z$ particle of the isotropic system has lowest energy whereas that of patchy particle is highest (see Fig.~\ref{fig:energy}). For patchy particle case, the orientation of the $Z$ particle at the center is incompatible with the six $\sigma$ neighbours. The energy of this hexagonal particle is expected higher than that of the $\sigma$ particle, and hence $\sigma$-dominent structures are expected at low density. The patchy particle system requires a fine tuning of density to assemble a DDQC, whereas the DDQC in isotropic particle system works even at low density condition (Fig.~\ref{fig:density}). Additional data can be found in Fig. \ref{fig:energy} and Fig.\ref{fig:density} in Appendix \ref{app:energy.density}.} 

\revC{Although the DDQCs in both systems consist of dodecagonal motifs, the network of the motifs is different. In isotropic DDQC (Fig.~\ref{fig:isotropic.rotation}), the motifs are arranged into the dense packing types as illustrated in Fig.~\ref{fig:motif.pack}(a,b,d). The network of the isotropic DDQC is more compacted so as the energy is minimised.} 

\section{Discussions and Conclusions}
We have investigated the formation of dodecagonal quasicrystals from the patchy particles \revA{with five-fold symmetry}. From a $Z$-rich hexagonal structure, \revA{fluctuations of positions of patchy particles generate intermediate structures} $D_1$, $D_2$. After the dodecagonal motif is formed, the motif is rapidly expanded and the $D_1, D_2$ are \revA{accumulated as a boundary between the cluster of dodecagonal motif} and the hexagonal structure. The DDQC contains many dodecagonal motifs packed in different ways. The formation of the DDQC is driven from the local fluctuations of the particles. The fluctuation is characterised by rotation of the first ring in the dodecagonal motif. After the DDQC is formed, the dodecagonal motif network changes \revA{when thermal fluctuation makes large displacements of the particles}. The change of the network is associated with rotation of the nearest neighbours of the centres of the motif.     

\revC{
Regarding the formation of the DDQC, the transformation from a $Z$-rich structure to a DDQC is observed in both patchy particle system and isotropic particle system. Superposition of a hexagonal lattice and a dodecagonal motif reveals the difference in the position of the first ring, which is manifested by a circular rotation of $15^{\circ}$ around the motif's centre. Such displacements of the particles on the first rings of the dodecagonal motifs are confirmed in our simulations. This maybe the reason why the DDQCs of patchy particles and isotropic particles both share same mechanism although the interactions are completely different. 
}

\revC{
The DDQC comprises different types of the packing of dodecagonal motifs (see Fig.\ref{fig:motif.pack}). These packing structure can be constructed from the network of centres of dodecagonal motifs. The packing is divided into several characteristic structures (Fig.\ref{fig:motif.pack}). In the obtained DDQCs, different types of the packing structures coexist, and they do not show periodic unit.
 We have observed that the DDQC continues to fluctuate in the sense that the relative positions of some particles change. Two types of displacement are considered. (i) In the case of individual displacement, the particles on the first ring of the dodecagonal motif quickly rotate and swap their positions. Despite the particle displacements, the whole dodecagonal motif remains intact. (ii) Collective displacement of particles can bring about the rearrangement of the DDQC network. Such changes appear at the border of different types of packing structures, i.e. some dodecagonal motifs are disintegrated and new motifs are generated. The displacement of the particle mostly occur at the first ring of the dodecagonal motifs, and at short lengthscale like (i). However such displacement can propagate to the neighbouring motifs, and the whole DDQC network eventually changes.} 


There is another type of DDQC made of isotropic particles \cite{engel_2010} where the the high-symmetry dodecagonal motif has five particles are on the inner ring (compared to six particles for the DDQC in our study) and twelve particles are on the second ring. Kinetics and dynamics of those DDQCs may be different from those in this study. We leave the analyses of such different DDQCs as a future study.   

The patchy particle DDQC in our study is in agreement with other works using five-fold symmetry patchy particle system, such as particle with five equally distributed patches on the equator \cite{vanderlinden_2012}. 
The $Y_{55}$ patchy particle in our study also has five-fold symmetry, but the number of patches and the interaction are different. This particle has ten patches in red and blue, where the interaction of red-red/blue-blue patches are assigned attractive and red-blue patches are repulsive. In the assembled DDQC, the orientation of the patchiness of the $\sigma$ particle and its neighbours (Fig. \ref{fig:local.structure}) is found similar to the one in ref.~\cite{vanderlinden_2012}. Moreover, the ratios of the $Z$ and $\sigma$ particles of the DDQC in the Monte-Carlo simulations in ref.~\cite{vanderlinden_2012} are about 0.07 and 0.8, respectively, which are comparable to our system in Fig.~\ref{fig:snapshot}.

\revC{The DDQC can be formed by bi-dispersed DNA system in simulation \cite{reinhardt_2017} or experimentation \cite{liu_2019a}. The two DNA tiles can be considered as 5- and 6-point-star particles. The self-assemblies of patchy particle DDQC at different densities in our study is somewhat analogical to that of bi-dispersed DNA systems \cite{liu_2019a}. When the ratio of the 5- to 6-point-star particle increases, the resulted structures change. Such a low ratio leads to $\sigma$ phase structures. Then medium ratio forms DDQC of which many dodecagonal motif are observed. At high ratio, the hexagonal structures dominate the dodecagonal motifs. Such changes are also observed in our study when the density of the patchy particle increases.}

\revC{We have investigated the formation and fluctuation of dodecagonal quasicrystals in patchy particle system and isotropic particle system. It may help to predict and design the two-dimensional dodecagonal quasicrystals in different soft-matter systems. The fluctuation of the DDQC may relate to the properties of the QC, e.g. heat capacity. An analogical study is still required for, for example, octagonal QC and decagonal QC to know how the other quasicrystals behave.}




\begin{acknowledgments}
The authors acknowledge the support from JSPS KAKENHI Grant number JP20K14437 to U.T.L., and JP20K03874 and JP20H05259 to N.Y.
\end{acknowledgments}


\appendix 
\section{Potential of patchy particle system}
\label{app:potential}
The potential for a pair of patchy particle is
\begin{equation} \label{eq:u}
V=V_{WCA}(r) + V_M(r)\Xi(\bm{\Omega})  ,
\end{equation}

The detail of the isotropic Week-Chandler-Anderson potential $V_{WCA}$ preventing the overlapping of particles, and the Morse potential $V_M$ in Eq.~\eqref{eq:u} is given as 
	\begin{equation}
	V_{WCA}= 
  	\begin{cases}
  	4\epsilon \left[ (\frac{2a}{r})^{12}- (\frac{2a}{r})^{6}+\frac{1}{4} \right]	, & r\leq 2a\sqrt[6]{2} \\
  	0 												    							, & r > 2a\sqrt[6]{2}
  	\end{cases} 
	\end{equation}

	\begin{equation}
	V_M= \epsilon M_d \left \{ \left[ 1-e^{\left( -\frac{r-r_{eq}}{M_r} \right)} \right]^2 -1 \right \} ,
	\end{equation}
where $\mathbf{r}=\mathbf{r}^{ij}=\mathbf{r}^j-\mathbf{r}^i$ is the distance vector between particle centres, $r=\left|\mathbf{r}\right|$, and $\hat{\mathbf{r}}=\mathbf{r}/r$, $\epsilon$ is the potential well depth, $r_{eq}$ is the Morse potential equilibrium position ($r_{eq}=1.878a$), $M_d=2.294a$, $M_r=a$ is the Morse potential depth and range, respectively \cite{delacruz-araujo_2016,lieu_2020,lieu_2022}.

The anisotropic interaction is calculated based on the mutual orientation of a pair of particle $i$ and $j$. Let the $\nvec^{(m)}_i$, $\nvec^{(m)}_j$, $m=1,2,3$ are local bases of particle $i$ and $j$, and $\rvec$ is the unit distance vector between particle center. The anisotropic interaction $\Xi_{lm}  \propto \{ \mathbf{C}_{(i)}^{(l,m)} \} \odot   \nabla_{\mathbf r}^{2l} \frac{1}{r}   \odot \{ \mathbf{C}_{(j)}^{(l,m)} \}$ estimates the angular dependence of a pair of particles $Y_{lm}$ as $\Xi_{lm}  \propto \{ \nvec_0^{l-m} \nvec_+^{m}\}_{(i)} \odot   \{\rvec^{2l} \}   \odot  \{ \nvec_0^{l-m} \nvec_+^{m}\}_{(j)} $, where $\nvec_0=\nvec^{(3)}$ and $\nvec_+=\frac{1}{\sqrt{2}}(\nvec^{(1)} + i \nvec^{(2)})$. For example,  $\Xi_{10}  \propto \{ \nvec_0 \}_{(i)} \odot   \{\rvec \rvec \}   \odot  \{ \nvec_0  \}_{(j)} $ for a pair of particles $Y_{10}$, and $\Xi_{20}  \propto \{ \nvec_0  \nvec_0\}_{(i)} \odot   \{\rvec \rvec \rvec \rvec \}   \odot  \{ \nvec_0 \nvec_0 \}_{(j)} $ for a pair of particles $Y_{20}$.  

\section{Isotropic particle system}  
\label{app:isotropic}
\begin{figure}[ht] 
  	\centering
   \includegraphics[width=0.35\textwidth]{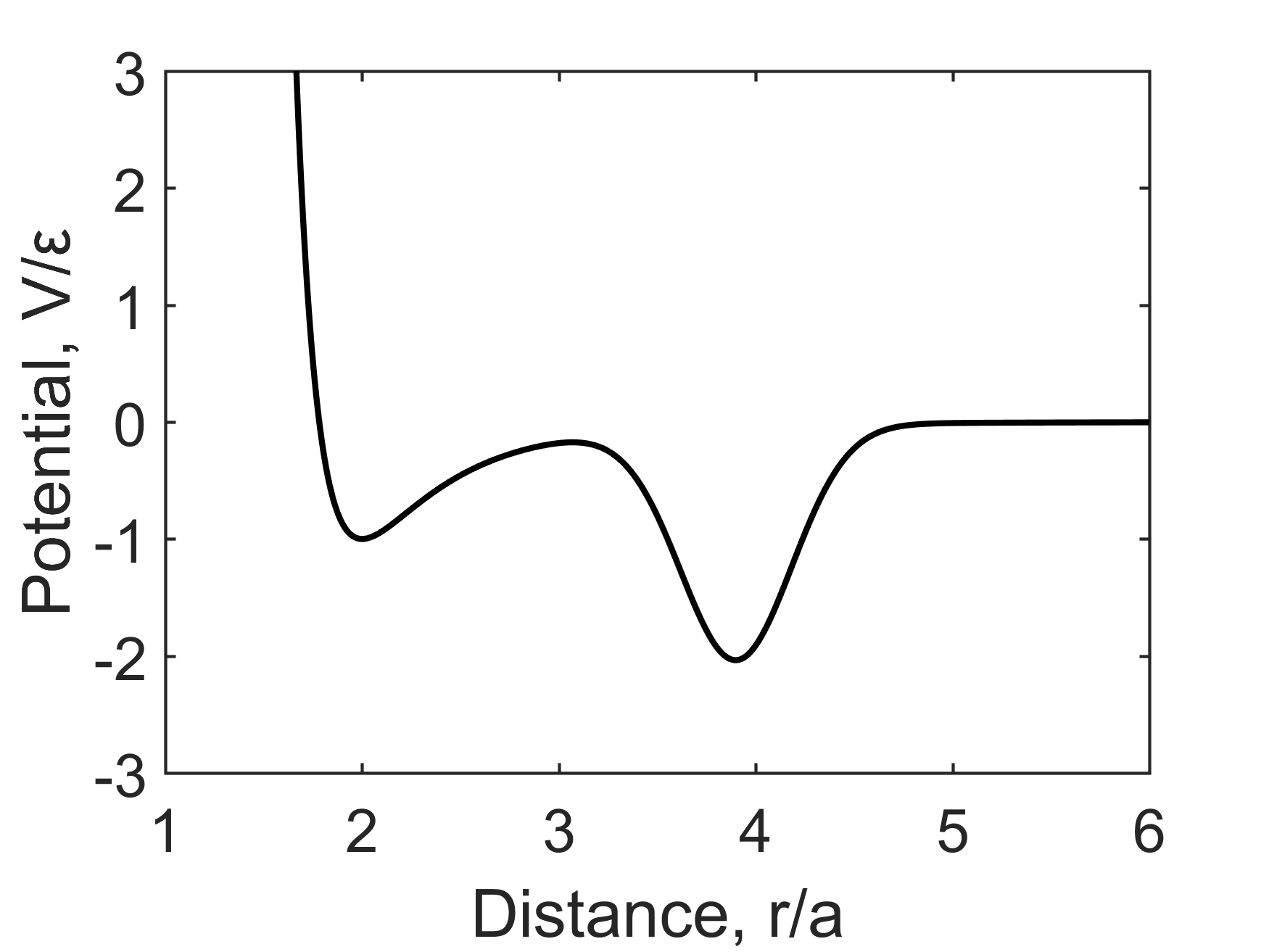}
   	\caption{Isotropic potential for the creation of DDQC.}
   	\label{fig:potential_iso}
	\end{figure}

The isotropic potential is a Lennard-Jones Gauss potential \cite{engel_2010,barkan_2014,gemeinhardt_2018,martinsons_2018} as follows 
	\begin{equation}
	V_{\text{LJG}}=\epsilon 	\left[ 	\left( 	\frac{2a}{r} \right)^{12} - 2 \left( \frac{2a}{r} \right)^{6} -\epsilon^{'} \exp \left(  \frac{(r-r_G)^2}{2 (2a)^2 \sigma^2}       \right)
									     	\right],
	\end{equation}
where the particle radius is $a$, the potential parameters $r_G$, $\epsilon^{'}$, $\sigma^2$ determine the position, depth, and width of the second well. The values of those parameters are $r_G=1.95\times2a$, $\epsilon^{'}=2.0$, $\sigma^2=0.02$. Figure \ref{fig:potential_iso} shows the potential.
	
The simulation parameters are almost identical to that of patchy particle system. The number of particles is $N=1024$. In the annealing simulations, the temperature decreases from $T=1.2$ to $T=0.4$ with an interval of $\Delta T=0.0125$. There are 100,000 steps at each value of temperature, making a total of $65\times100,000$ steps.   

\section{Determination of neighbouring rotation}  
\label{app:rotation}
\begin{figure}[ht]
	\centering
	\includegraphics[width=0.45\linewidth]{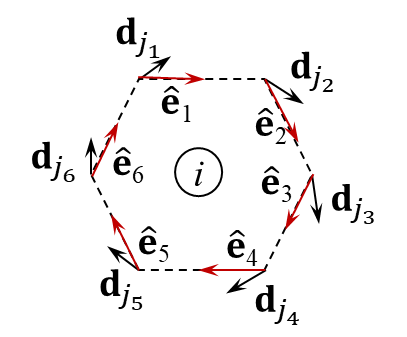}
 	\caption{Calculation of local rotation}
  	\label{fig:rotation.cal}
   \end{figure}

The particle configuration is known at every time step. Let $\mathbf{r}(t_1)$, $\mathbf{r}(t_2)$ be the particle position at time $t_1$, $t_2$, respectively, and $\mathbf{d}=\mathbf{r}(t_2)-\mathbf{r}(t_1)$ the particle displacement from $t_1$ to $t_2$. Consider that the particle $i$ has $N_i$ nearest neighbours $\{j_1, j_2, ...,j_{N_i}\}$ in clockwise order. At $t_1$, we define $\hat{\mathbf{e}}_{j_k}$ the unit vector from $j_k$ to $j_(k+1)$ of the polygon made of $\{j_1, j_2, ...,j_{N_i}\}$. The displacement of the neighbours particle is determined from $t_1$ to $t_2$, for example, $\mathbf{d}_{j_1}$ is the displacement of the particle $j_1$ (Fig. \ref{fig:rotation.cal}). The rotation of the neighbours of particle $i$ is defined as
\begin{equation}
\omega_i=\frac{1}{N_i}\sum_{j} \mathbf{d}_{j_k} . \hat{\mathbf{e}}_{j_k}.
\end{equation}
Note that the relative displacement to the center particle is considered $\mathbf{d} \mapsto \mathbf{d}-\mathbf{d}_i$.

\section{\revC{Energy of local structure and density dependence}}  
\label{app:energy.density}
\begin{figure}[ht] 
  	\centering
  	\includegraphics[width=0.35\textwidth, trim=10mm 115mm 110mm 5mm,clip]{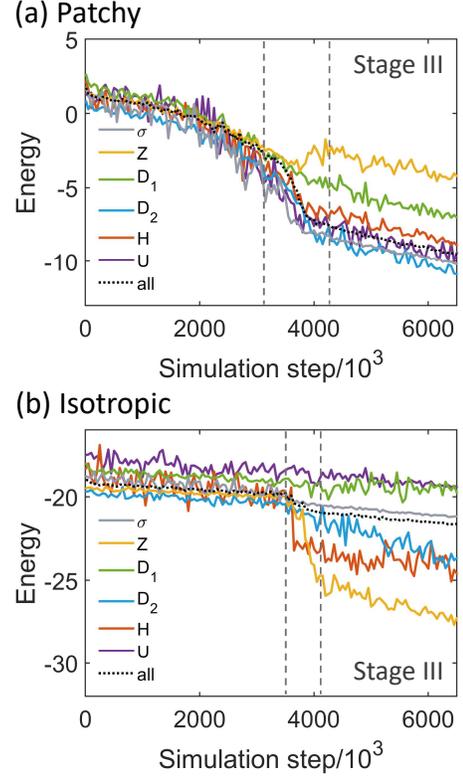}   	   	
   	\caption{\revC{Energy of local structures during annealing simulation from $T=1.2$ to $T=0.4$ for (a) patchy particle system  and (b) isotropic particle system. The colours of local structures are given in Fig.~\ref{fig:local.structure}, and the black dotted line shows the average energy of all particles. The vertical dash lines define the three stages of the growth of the DDQC (see Fig.~\ref{fig:snapshot}).}}   
   	\label{fig:energy}
	\end{figure} 
	
	\begin{table}[h]
	\small
\centering
\setlength{\tabcolsep}{4pt}
\begin{tabular}{*{6}{l}} 
 $\#$	& $\rho_{\text{a}}$ 	&		 $T_{\text{max}}$ &		 $T_{\text{min}}$  &		 $\Delta T$		&Total simulation steps \\ 
 	\hline
 	\multicolumn{4}{l}{Patchy particles} \\
	(a)&  0.45 & 1.0 & 0.2	&0.0125 & $13000\times10^3$ \\  
	(b)& 0.84 & 1.2 & 0.4	&0.0125 & $6500\times10^3$ \\  
	\multicolumn{4}{l}{Isotropic particles} \\
	(c)& 0.30 & 1.2 & 0.2	&0.0125 & $8500\times10^3$ \\  
	(d)& 0.87 & 1.2 & 0.2	&0.0125 & $8500\times10^3$ \\  
\end{tabular}
\caption{Parameters of annealing simulations at different densities. The results are shown in Fig.\ref{fig:density}}
\label{table:1}
\end{table}

\begin{figure}[ht] 
  	\centering
  	\includegraphics[width=0.45\textwidth, trim=12mm 77mm 90mm 2mm,clip]{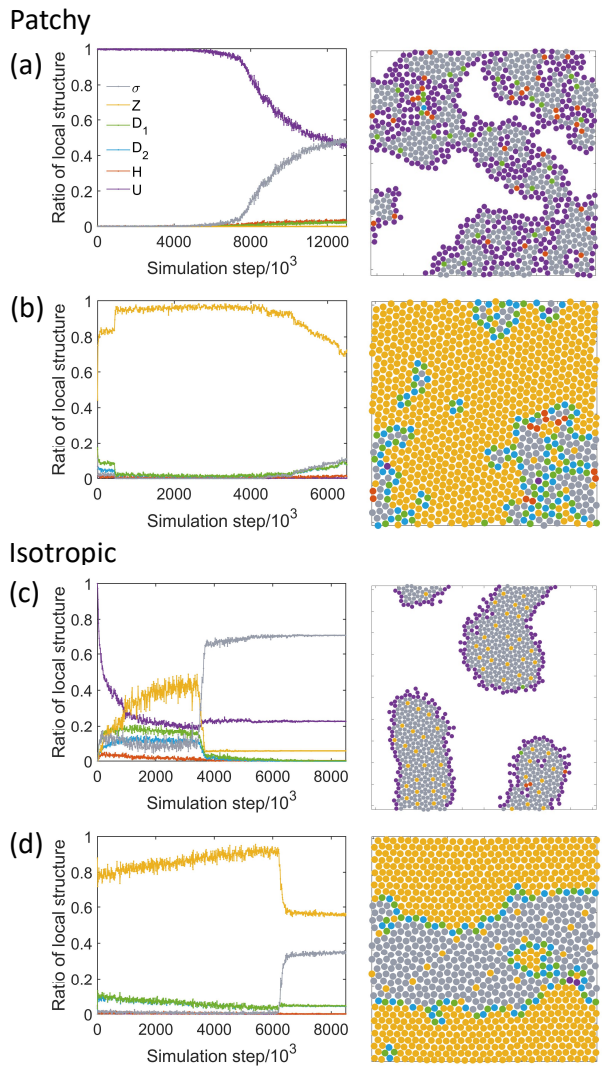}
  	\caption{Changes in the ratio of local structures of patchy particles (a-b) and isotropic particles (c-d) self-assemblies at low and high values of area fractions. The snapshots are taken at the last steps of the annealing simulations. The area fractions $\rho_{\text{a}}$ of the simulations (a-d) are 0.45, 0.84, 0.30, 0.87, respectively. Details of simulation parameters are given in Table \ref{table:1}.}
   	\label{fig:density}
	\end{figure} 

We compare the energy of each local structure during the growth of the DDQC assembled from the patchy particles and isotropic particles. In each snapshot, for example, the energy of the $Z$ particle is the mean energy of all $Z$ particles. The data are calculated from the annealing simulations of Fig.~\ref{fig:snapshot} and Fig.~S1 in Supplementary Information. 
	
The energy of each local structure during annealing is given in Fig. \ref{fig:energy}. The DDQC is steadily formed when the ratio of the local structure is in stage III. Regarding the local structure of DDQC, the lowest energy is the $\sigma$ particle in patchy DDQC, while that is $Z$ particle in isotropic DDQC. This is due to the difference in two particle systems. The five-fold patchy particle prefers five nearest neighbours. Therefore, the local structure $\sigma$ is expected to have lowest energy. Then self-assembly of patchy particles at low densities result in $\sigma$-rich structure instead of DDQC. In contrast, the $Z$ particle in isotropic DDQC has lowest energy, and therefore the DDQC structure is formed at low densities. The decrease of energy with time when the DDQC reaches stable stage III is mainly due to lowering temperature in annealing. As a result, the thermal fluctuation is reduced and the energy decreases. 	
	
We investigate the effect of density on the formation of DDQC by patchy particles and isotropic particles. Annealing simulations are conducted with varying the area fractions of the particles. The simulation parameters and results are given in Table \ref{table:1} and Fig.~\ref{fig:density}. 

There are different phases depending on the area fraction. For the patchy particle system, the suitable density for the assembly of DDQC is limited around $ 0.69 \lesssim \rho_{\text{a}} \lesssim 0.81$. At low density $\rho_{\text{a}} < 0.69$, the structure contains clusters of $\sigma$ particles and undefined particles at the interfaces. The dodecagonal motif is hardly observed (Fig.\ref{fig:density}(a)). At high density $\rho >0.81$ the hexagonal structure is dominant (Fig.~\ref{fig:density}(b)). 

For the isotropic particle system, the DDQC more easily as $\rho_{\text{a}} \lesssim 0.84$. Even at the low density, clusters of the dodocagonal motif are clearly observed inside the interfaces (Fig.~\ref{fig:density}(c)). At high density $\rho >0.84$ the hexagonal structure is also dominant (Fig.~\ref{fig:density}(d)).





\bibliography{MyLibrary} 
\bibliographystyle{unsrt}
\end{document}


\title{Formation and Fluctuation of Two-dimensional Dodecagonal Quasicrystal \\
SUPPLEMENTARY INFORMATION}

\author{Uyen Tu Lieu}
\email{uyen.lieu@aist.go.jp}
\affiliation{Mathematics for Advanced Materials-OIL, AIST, 2-1-1 Katahira, Aoba, 980-8577 Sendai, Japan}

\author{Natsuhiko Yoshinaga}
\email{yoshinaga@tohoku.ac.jp}
\affiliation{Mathematics for Advanced Materials-OIL, AIST, 2-1-1 Katahira, Aoba, 980-8577 Sendai, Japan}
\affiliation{Advanced Institute for Materials Research (AIMR), Tohoku University, 2-1-1 Katahira, Aoba, 980-8577 Sendai, Japan}


\begin{abstract}
\end{abstract}

\maketitle


	\begin{figure*}[ht] 
   	\centering 
 		\includegraphics[width=0.9\textwidth, trim=10mm 145mm 10mm 10mm,clip]{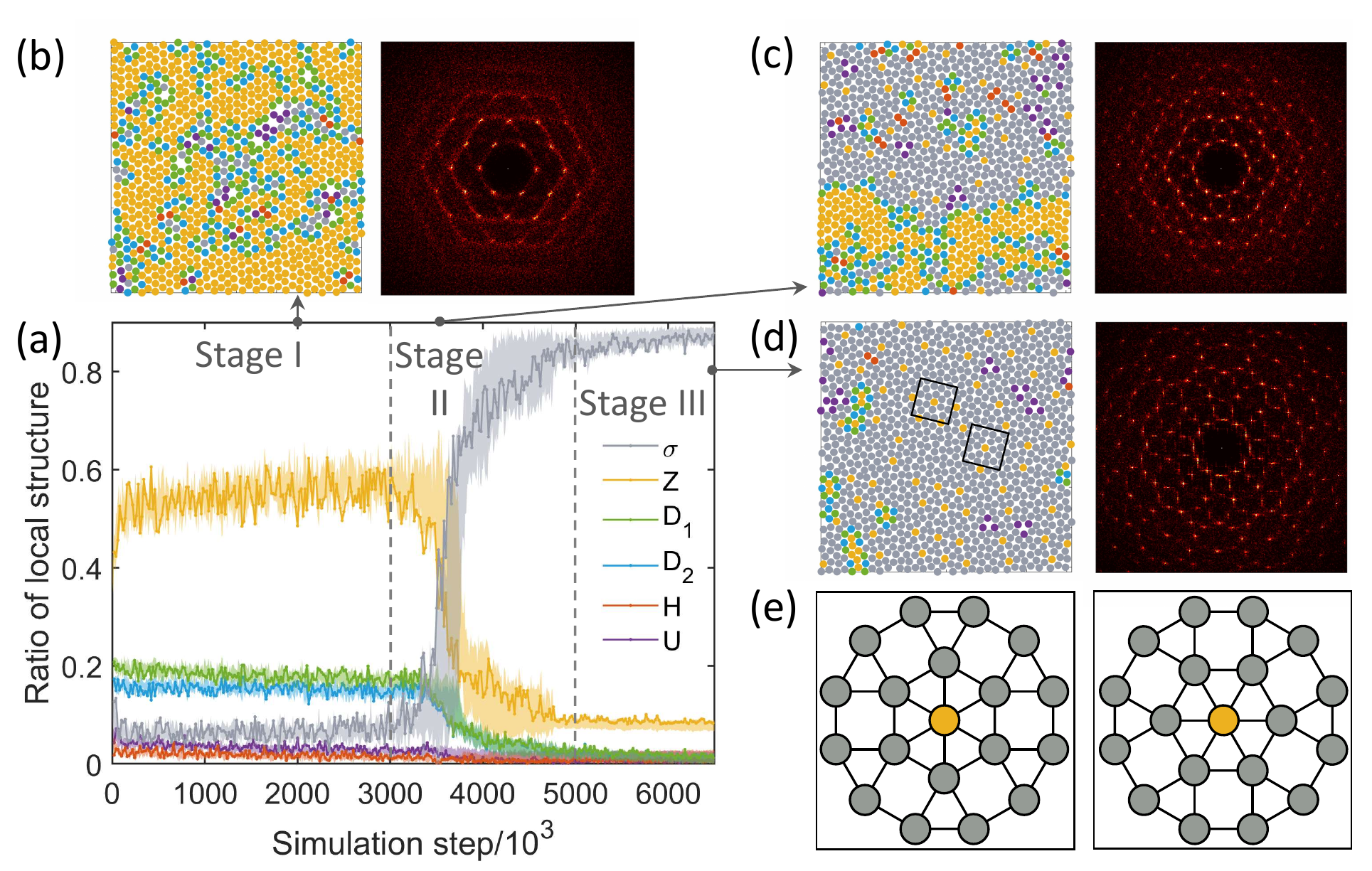}        
   	\caption{Change in local structures of isotropic particles during annealing simulations at $\rho_\text{a}=0.825$. (a) Ratio of the local structure in time; the line represents the data of a simulation, the shade area corresponds to the 95$\%$ confidence intervals of 5 independent simulations. The growth of the dodecagonal quasicrystal is divided into three stages. (b-d) The snapshots and their Fourier transformations. (e) Two kinds of dodecagonal motif as shown in the two boxes in (d), they are interchangeable by rotating the particles on the first ring $\pm30^\circ$. 
   	}
   	\label{fig:snapshot}
	\end{figure*}

\begin{figure}[ht] 
   	\centering 	    
 		\includegraphics[width=0.45\textwidth, trim=10mm 125mm 105mm 5mm,clip]{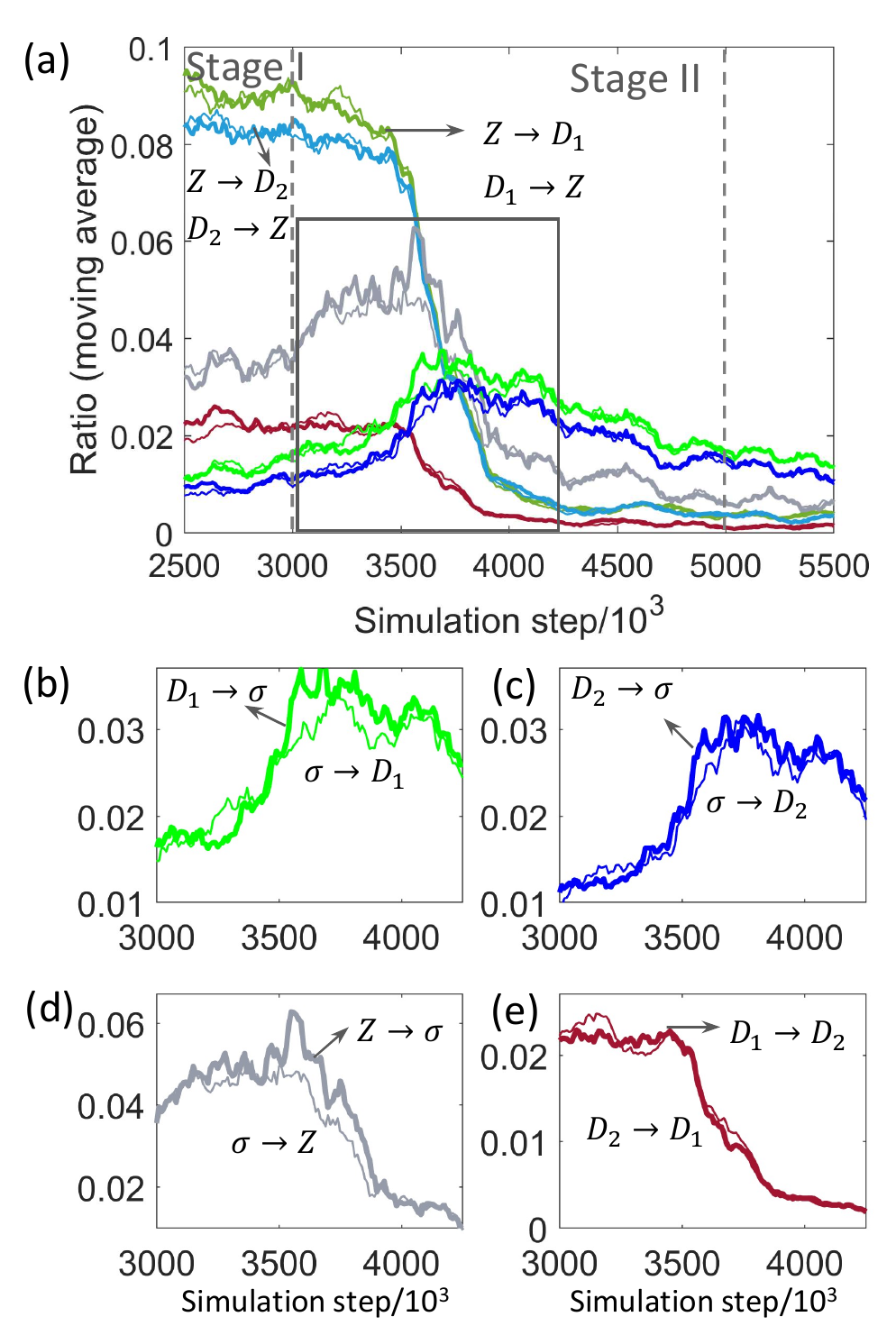}
   	\caption{The interchange between the local structures during the growth of DDQC made of isotropic particles at $\rho_\text{a}=0.825$ by decomposition of the ``in'' and ``out'' amount of selected pairs of local structures (a). (b-e) The corresponding data inside the box in (a). The graph show moving average of 15 data points.}
   	\label{fig:interchange}
	\end{figure}		

\begin{figure}[ht] 
   \centering   
   \includegraphics[width=0.40\textwidth, trim=30mm 30mm 40mm 5mm,clip]{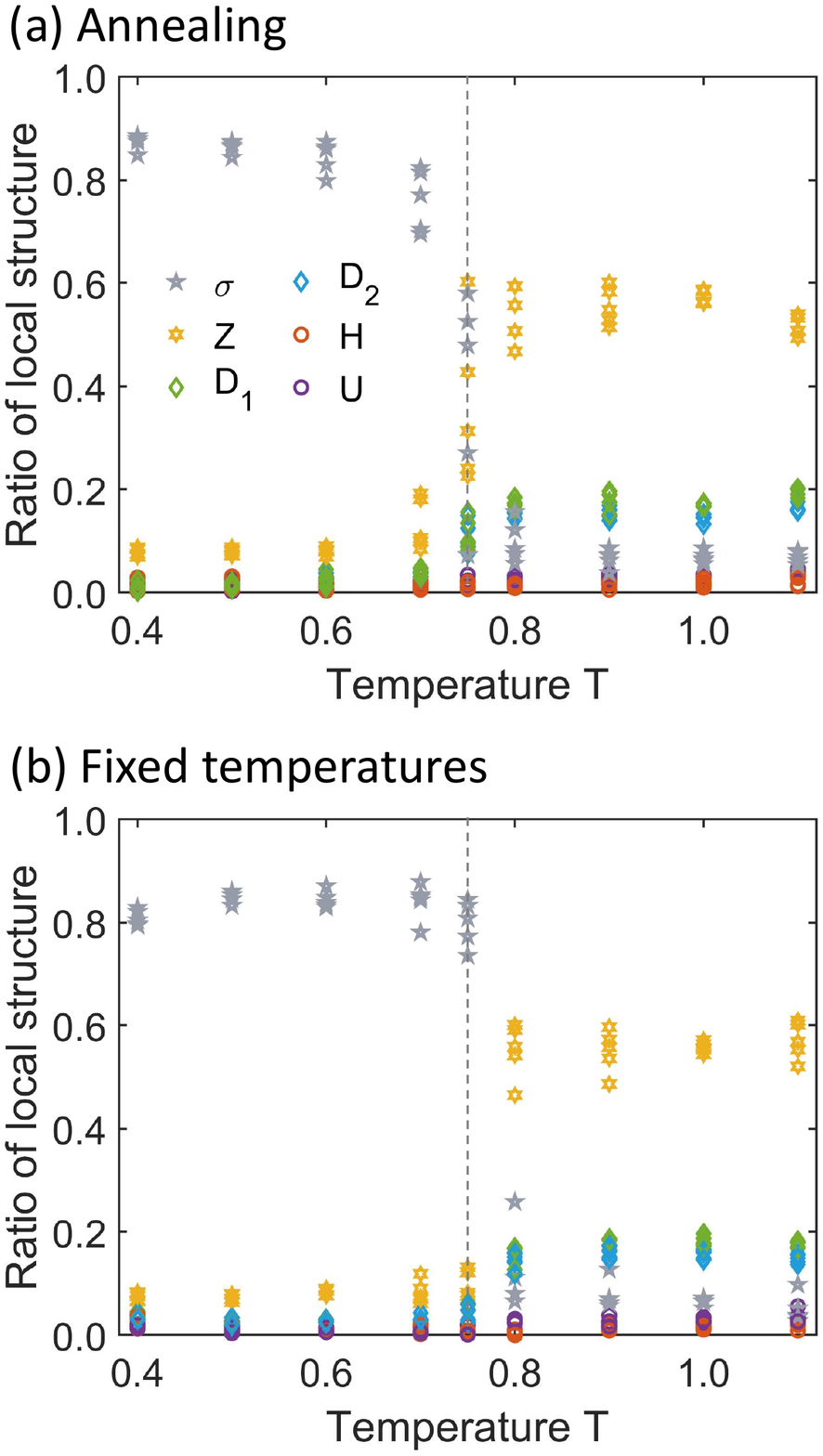}
   \caption{Dependence of ratio of local structure of isotropic particles on temperature in (a) annealing simulations, (b) fixed temperature simulations at $\rho_{\text{a}}=0.825$. The dashed lines approximate the critical temperature $T^{*}$.}
   \label{fig:ratio.vs.temp}
   \end{figure}
